\documentclass[aip,graphicx]{revtex4-1}

\usepackage{mathrsfs,amsmath}
\usepackage{amsfonts}
\usepackage{amssymb}
\usepackage{graphicx}
\usepackage{epstopdf}
\usepackage{float}
\usepackage{romannum}
\usepackage{xcolor}

\draft 

\begin{document}


\title{Linear unstable whistler eigenmodes excited by a finite electron beam} 



\author{Xin An}
\email[]{xinan@atmos.ucla.edu}
\affiliation{Department of Atmospheric and Oceanic Sciences, University of California, Los Angeles, California, 90095, USA}

\author{Jacob Bortnik}
\affiliation{Department of Atmospheric and Oceanic Sciences, University of California, Los Angeles, California, 90095, USA}

\author{Bart Van Compernolle}
\affiliation{Department of Physics and Astronomy, University of California, Los Angeles, California, 90095, USA}


\date{\today}

\begin{abstract}
Electron beam-generated whistler waves are widely found in the Earth's space plasma environment and are intricately involved in a number of phenomena. Here we study the linear growth of whistler eigenmodes excited by a finite gyrating electron beam, to facilitate the interpretation of relevant experiments on beam-generated whistler waves in the Large Plasma Device at UCLA. A linear instability analysis for an infinite gyrating beam is first performed. It is shown that whistler waves are excited through a combination of cyclotron resonance, Landau resonance and anomalous cyclotron resonance, consistent with our experimental results. By matching the whistler eigenmodes inside and outside the beam at the boundary, a linear growth rate is obtained for each wave mode and the corresponding mode structure is constructed. These eigenmodes peak near the beam boundary, leak out of the beam region and decay to zero far away from the beam.
\end{abstract}

\pacs{}

\maketitle 

\section{Introduction and problem setup}
Electron beam-generated whistler waves play an important role in a number of space plasma settings. To name just a few examples, intense electron beams form in the separatrix region of magnetic reconnection and excite whistler waves \cite{zhang1999whistler, deng2001rapid, wei2007cluster, fujimoto2014wave, huang2016two}, which were thought to mediate the rapid energy release involved in magnetic reconnection \cite{mandt1994transition, shay1999scaling, birn2001geospace}. Electron beams accelerated by solar flares excite Langmuir waves, which further generate type \Romannum{3} radio bursts through nonlinear wave-wave interactions \cite{ginzburg1959mechanisms, robinson1997nonlinear}. In particular, electron beam-generated whistler waves interact with Langmuir waves and drive cyclic Langmuir collapse in the nonlinear stage of the electron two-stream instability \cite{che2017electron}, which was considered to be the key step in reconciling the huge difference in time scales between the short linear saturation of electron two-stream instability and long duration of type \Romannum{3} radio bursts. In the Earth's quasiperpendicular bow shock, electron beam-generated whistler waves were seen to contribute to the electron heating throughout the shock transition layer \cite{tokar1985propagation, chen2018electron}. Whistler-mode auroral hiss can be generated by electron beams in the ionosphere \cite{maggs1978electrostatic} and potentially leak to the ground. In the Earth's outer radiation belt, the lower-band oblique whistler-mode chorus waves are suggested to be generated by electron beams \cite{mourenas2015very, li2016unraveling}, though the origin of these electron beams seems unclear at this time. In addition, beam-plasma interactions have also been extensively studied in active experiments in space \cite{winckler1980application} and controlled experiments in laboratory \cite{stenzel1977observation, krafft1994whistler, starodubtsev1999resonant} for various purposes.

By injecting a circular electron beam into a cold background plasma, we performed a series of laboratory experiments \cite{van2015excitation, an2016resonant, van2016laboratory} to excite whistler waves in the Large Plasma Device (LAPD) \cite{gekelman2016upgraded} at the University of California, Los Angeles (UCLA), supported by particle-in-cell simulations \cite{an2017electrostatic}. Of relevance to the present study, one of our experiments \cite{an2016resonant} identified the resonant mode structures of hiss-like emissions excited by a gyrating electron beam with finite size, and observationally demonstrated that these emissions were excited through a combination of cyclotron resonance, Landau resonance and anomalous cyclotron resonance. On the theoretical side, Bell and Buneman \cite{bell1964plasma} had investigated the linear growth of parallel-propagating whistler waves excited by an infinite gyrating electron beam through cyclotron resonance only which could account for a portion of the instability but not the entire excitation process. The resonant growth or damping of whistler waves had been treated in detail by Kennel \cite{kennel1966low} for a plasma consisting of a dense, cold background component and an energetic nonthermal component, in which whistler waves can propagate at an arbitrary angle to the magnetic field. But the frequency of whistler waves $\omega$ is restricted to the range $\Omega_i \ll \omega \ll \Omega_e$ in Kennel's treatment ($\Omega_i$ and $\Omega_e$ are the ion and electron cyclotron frequency, respectively). The general dielectric tensor for a gyrating electron beam was obtained in Ref.~\onlinecite{akhiezer2017plasma}, in which the whistler instability was discussed briefly. Elliot \cite{elliott1975ducting} investigated the ducting of various unstable wave modes in a thin field-aligned electron beam. Dungey and Strangeway \cite{dungey1976instability} further developed the work of Elliot with the emphasis on the stability of different wave modes.

To address the whistler instabilities that are pertinent to our experiment, here we consider a finite electron beam in a slab geometry which nevertheless allows us to capture the essential ingredients of the experiment but does not unnecessarily complicate the solution mathematically. The background magnetic field $\mathbf{B}_0$ is taken to be along the $z$ axis. The density of the electron beam is assumed to have a simple top-hat profile in the perpendicular ($x$) direction, namely
\begin{eqnarray}
n_b(x) = \begin{cases}
n_b,\, |x| < a \\
0,\, \mbox{otherwise}
\end{cases} ,
\end{eqnarray}
where $a$ is the half width of the beam. The background electrons are assumed to be cold and uniform, with a density of $n_0$. The background ions are taken to be cold and are distributed in such a way as to maintain the neutrality of the system. The frequency range considered in the present analysis here is much higher than the lower hybrid frequency so that ions can be treated as a fixed background. The unperturbed electron beam distribution function is
\begin{eqnarray}
f_{0b}(v_\parallel, v_\perp) = \frac{n_b}{2\pi v_\perp} \delta(v_\perp - v_{\perp 0}) \delta(v_z - u) .
\end{eqnarray}
The wave dispersion relation inside and outside the beam, respectively, can be treated as a uniform medium. The linear instability of an electron beam of a finite size differs from that of an infinite homogeneous beam, in that the unstable waves spend a limited amount of time inside the beam for amplification and eventually propagate out of the beam region. Such a finite electron beam will likely lead to a decrease of the linear wave growth rate compared to an infinite electron beam. Note that the size of the electron beam is only a few times the gyro-radius of the beam electrons and is comparable to the wavelength of whistler waves consistent with our experimental setup at the LAPD. This ordering of the length scales violates the assumption behind the ray tracing method and hence it cannot be applied in the present situation. Corresponding to each complex wave frequency (eigenvalue), the wave has a certain mode structure (eigenmode) that may either leak out of the beam region or be confined to the beam region. In section \ref{sec:linear_infinite_beam}, a linear instability analysis is performed for an infinite electron beam. In section \ref{sec:match_boundary_condition}, the eigenmode solutions are matched at the boundary and the results for a finite electron beam are presented. In section \ref{sec:experiment}, the experimental results are quantitatively interpreted with the linear instability analysis. The work is summarized and further discussed in section \ref{sec:sum-discuss}.

\section{Linear instability analysis for an infinite electron beam}\label{sec:linear_infinite_beam}
To begin with, the linearized Vlasov equation can be written as:
\begin{eqnarray} \label{eqn:linear_vlasov}
\frac{\partial \hat{f}_b}{\partial t} + \mathbf{v} \cdot \boldsymbol{\nabla} \hat{f}_b -\frac{e}{m} \left( \frac{\mathbf{v}\times \mathbf{B}_0}{c} \right) \cdot \frac{\partial \hat{f}_b}{\partial \mathbf{v}} = \frac{e}{m} \left( \hat{\mathbf{E}} + \frac{\mathbf{v}\times \hat{\mathbf{B}}}{c} \right) \cdot \frac{\partial f_{0b}}{\partial \mathbf{v}} ,
\end{eqnarray}
where $\hat{f}_b$ is the perturbed distribution function and $\hat{\mathbf{E}}$, $\hat{\mathbf{B}}$ are the perturbed fields. $e$ is the elementary charge and $m$ is the electron mass. We consider perturbations of the form
\begin{eqnarray}
\hat{\mathbf{E}} &=& \tilde{\mathbf{E}} \exp(-i\omega t + i k_x x + i k_z z) , \\
\hat{\mathbf{B}} &=& \tilde{\mathbf{B}} \exp(-i\omega t + i k_x x + i k_z z) .
\end{eqnarray}
Equation \eqref{eqn:linear_vlasov} can be integrated along its characteristics, i.e., the unperturbed helical orbits of the electrons. This integral can be calculated as
\begin{eqnarray} \label{deltaf}
\hat{f}_b &=& \frac{e}{m} \int_{-\infty}^{t} d t^{\prime} \exp(-i\omega t^\prime + i k_x x^\prime + i k_z z^\prime) \tilde{S} .
\end{eqnarray}
Here the integral kernel $\tilde{S}$ is
\begin{eqnarray} \label{int_kernel}
\tilde{S} &=& \left( \tilde{\mathbf{E}} + \frac{\mathbf{v}^\prime \times \tilde{\mathbf{B}}}{c} \right) \cdot \frac{\partial f_{0b}(\mathbf{v}^\prime)}{\partial \mathbf{v}^\prime} .
\end{eqnarray}
$x^\prime$ and $z^\prime$ represent the position at time $t^\prime$ along the particle orbit. $\mathbf{v}^\prime$ represents the velocity at time $t^\prime$ along the particle orbit. The detailed expressions of $x^\prime$, $z^\prime$ and $\mathbf{v}^\prime$ can be found in the supplemental materials. The perturbed beam current can be calculated from the velocity moments of the perturbed beam distribution $\hat{f}_b$ as
\begin{eqnarray}\label{jb}
\hat{\mathbf{j}}_b = -e \int_0^\infty 2\pi v_\perp d v_\perp \int_{-\infty}^{\infty} d v_z \langle \mathbf{v} \hat{f}_b \rangle_\phi .
\end{eqnarray}
Note that $\hat{\mathbf{j}}_b$ has the form $\hat{\mathbf{j}}_b = \tilde{\mathbf{j}}_b \exp(- i \omega t + i k_x x + i k_z z)$. In fact, $\tilde{\mathbf{j}}_b$ can be expressed as a linear superposition of $\tilde{E}_x(x)$, $\tilde{E}_y(x)$ and $\tilde{E}_z(x)$, with the coefficients being integrals of gradients over velocity space, namely
\begin{eqnarray}\label{jb_lin_supp}
\begin{split}
\frac{4 \pi i}{\omega} \tilde{\mathbf{j}}_{b} = \boldsymbol{\chi}_b \cdot \tilde{\mathbf{E}} .
\end{split}
\end{eqnarray}
Here $\boldsymbol{\chi}_b$ is the susceptibility tensor for the electron beam, which is derived in detail in the supplemental material. The susceptibility tensor is summarized here as the following
\begin{equation}
\chi_{xx} = - \frac{\omega_{pb}^2}{\omega^2} - \frac{\omega_{pb}^2}{\omega^2} \sum\limits_{n=-\infty}^{\infty} \left[ \left( \frac{2 n^2}{\lambda} J_n J_n^\prime \right) \frac{n \Omega_e}{\omega - k_z u - n\Omega_e} + \left( J_n^2 \cot^2 \theta \right) \frac{n^2 \Omega_e^2}{(\omega - k_z u - n \Omega_e)^2} \right] ,
\end{equation}
\begin{equation}
\chi_{yy} = - \frac{\omega_{pb}^2}{\omega^2} - \frac{\omega_{pb}^2}{\omega^2} \sum\limits_{n=-\infty}^{\infty} \left[ \left( \frac{1}{\lambda}(\lambda^2 (J_n^\prime)^2)^\prime \right) \frac{n \Omega_e}{\omega - k_z u - n\Omega_e} + \left( \lambda^2 (J_n^\prime)^2 \cot^2 \theta \right) \frac{\Omega_e^2}{(\omega - k_z u - n \Omega_e)^2} \right] ,
\end{equation}
\begin{equation}
\chi_{zz} = - \frac{\omega_{pb}^2}{\omega^2} \tan^2 \theta - \frac{\omega_{pb}^2}{\omega^2} \sum\limits_{n=-\infty}^{\infty} \left[ \left(\frac{2 n}{\lambda} J_n J_n^\prime \tan^2 \theta \right) \frac{(\omega - n \Omega_e)^2}{\Omega_e (\omega - k_z u - n \Omega_e)} + J_n^2 \frac{(\omega - n\Omega_e)^2}{(\omega - k_z u - n \Omega_e)^2} \right] ,
\end{equation}
\begin{equation}
\chi_{xy} = i \frac{\omega_{pb}^2}{\omega^2} \sum\limits_{n=-\infty}^{\infty} \left[ \left( \frac{n}{\lambda} (\lambda J_n J_n^\prime)^\prime \right) \frac{n \Omega_e}{\omega - k_z u - n \Omega_e} + \left( n \lambda J_n J_n^\prime \cot^2 \theta \right) \frac{\Omega_e^2}{(\omega - k_z u - n \Omega_e)^2} \right] ,
\end{equation}
\begin{equation}
\chi_{xz} = \frac{\omega_{pb}^2}{\omega^2} \tan \theta - \frac{\omega_{pb}^2}{\omega^2} \sum\limits_{n=-\infty}^{\infty} \left[ \left( \frac{2 n^2}{\lambda} J_n J_n^\prime \tan \theta \right) \frac{\omega - n \Omega_e}{\omega - k_z u - n \Omega_e} + \left( n J_n^2 \cot \theta \right) \frac{\Omega_e (\omega - n\Omega_e)}{(\omega - k_z u - n \Omega_e)^2} \right] ,
\end{equation}

\begin{equation}
\chi_{yz} = -i \frac{\omega_{pb}^2}{\omega^2} \sum\limits_{n=-\infty}^{\infty} \left[ \left( \frac{n}{\lambda} (\lambda J_n J_n^\prime)^\prime \tan\theta \right) \frac{\omega - n \Omega_e}{\omega - k_z v_z - n\Omega_e} + \left( \lambda J_n J_n^\prime \cot\theta \right) \frac{\Omega_e (\omega - n \Omega_e)}{(\omega - k_z u - n \Omega_e)^2} \right] ,
\end{equation}
\begin{equation}
\chi_{yx} = - \chi_{xy} ,
\end{equation}
\begin{equation}
\chi_{zx} = \chi_{xz} ,
\end{equation}
\begin{equation}
\chi_{zy} = - \chi_{yz} .
\end{equation}
Here $\omega_{pb}^2 = 4\pi n_b e^2 /m_e$. $\theta$ is the angle between the background magnetic field $\mathbf{B}_0$ and the wave vector $\mathbf{k}$. $J_n$ is the Bessel function of the first kind of $n$-th order and its argument is essentially the wave number normalized by the gyro-radius of the beam electron, given by $\lambda = k_x v_{\perp 0} / \Omega_e$. Note that the susceptibility tensor $\boldsymbol{\chi}_b$ is a Hermitian matrix for any real $\omega$. $\boldsymbol{\chi}_b$ is also proportional to the beam density, i.e., $\boldsymbol{\chi}_b \propto n_b$. Since the beam is tenuous, $\boldsymbol{\chi}_b$ is significant only when the residual of $\omega - k_z u - n \Omega_e$ is small. The singularity at the resonance is avoided by navigating around it in the complex plane of the wave frequency $\omega$.

\begin{figure}[h]
\centering
\includegraphics[width=0.675\textwidth]{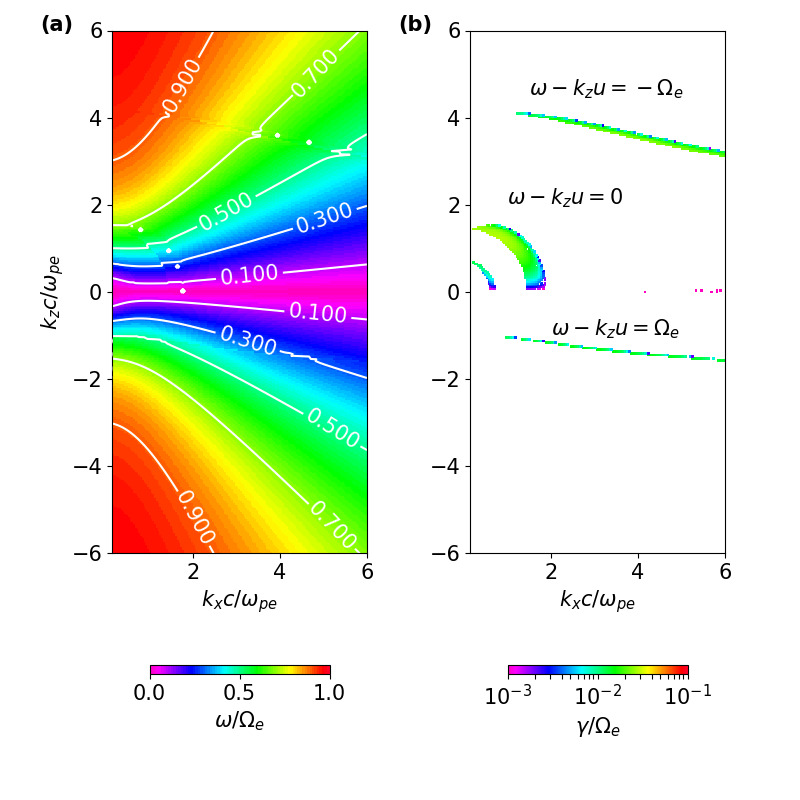}
\caption{(a) The real wave frequency as a function of wave number $k_x$ and $k_z$. (b) The corresponding wave growth rate in wave number space. All the wave numbers are normalized to the electron inertial length $c/\omega_{pe}$. All the frequencies are normalized to the electron cyclotron frequency $\Omega_e$.}
\label{fig-uniform-beam}
\end{figure}

Combining Faraday's and Ampere's Law, one obtains
\begin{eqnarray}\label{ampere}
\frac{c^2}{\omega^2}\mathbf{k} \times \mathbf{k} \times \tilde{\mathbf{E}} = - \left( \frac{4 \pi i}{\omega} \tilde{\mathbf{j}}_b + \frac{4 \pi i}{\omega} \tilde{\mathbf{j}}_c + \tilde{\mathbf{E}} \right) ,
\end{eqnarray}
where $\tilde{\mathbf{j}}_b$ and $\tilde{\mathbf{j}}_c$ are perturbed plasma currents generated by beam electrons and cold background electrons, respectively. Making use of the cold plasma dielectric tensor, one can combine the currents generated by the background electrons $\hat{\mathbf{j}}_c$ and the displacement current as
\begin{eqnarray}
\frac{4 \pi i}{\omega} \hat{\mathbf{j}}_c + \hat{\mathbf{E}} = \boldsymbol{\epsilon}_c \cdot \mathbf{E} ,
\end{eqnarray}
where $\boldsymbol{\epsilon}_c$ is the well-known cold plasma dielectric tensor \citep{stix1962wave}, written as
\begin{eqnarray}\label{eqn:cold-plasma-dielectric}
\boldsymbol{\epsilon}_c = \begin{pmatrix}
\epsilon_\perp &  -i\epsilon_H   &  0\\
i\epsilon_H    &  \epsilon_\perp &  0\\
0              &  0              &  \epsilon_\parallel
\end{pmatrix}.
\end{eqnarray}
Combining equations \eqref{ampere} - \eqref{eqn:cold-plasma-dielectric} leads to $\mathbf{M}_b \cdot \tilde{\mathbf{E}} = \boldsymbol{0}$ with the dispersion matrix having the form
\begin{eqnarray}
\begin{split}
\mathbf{M}_b =& \frac{c^2}{\omega^2}\mathbf{kk} - \frac{k^2 c^2}{\omega^2} \mathbf{I}_3 + \boldsymbol{\epsilon}_c + \boldsymbol{\chi}_b \\
=& \begin{pmatrix}
\epsilon_\perp + \chi_{xx} - \frac{k_z^2 c^2}{\omega^2}  &  -i \epsilon_H + \chi_{xy}  &  \chi_{xz} + \frac{k_x k_z c^2}{\omega^2} \\
i \epsilon_H + \chi_{yx}  &  \epsilon_\perp + \chi_{yy} - \frac{(k_x^2 + k_z^2) c^2}{\omega^2}  &  \chi_{yz} \\
\chi_{zx} + \frac{k_x k_z c^2}{\omega^2}  &  \chi_{zy}  &  \epsilon_\parallel + \chi_{zz} - \frac{k_x^2 c^2}{\omega^2}
\end{pmatrix}.
\end{split}
\end{eqnarray}
For nontrivial solution, we write $\det(\mathbf{M}_b) = 0$, which gives the dispersion relation of the beam-plasma system. For a given point in wave number space $(k_x, k_z)$, a root finding procedure is implemented to solve the dispersion relation for a complex wave frequency. The input background parameters for the numerical calculation are $\omega_{pe}/\Omega_e = 5.0$, $n_b/n_0 = 0.0016$. The input beam parameters for a $3$\,keV beam with $30$ degree pitch angle are $u/c = 0.094$ and $v_{\perp 0}/c = 0.054$. These numbers are motivated by the experimental parameters of the LAPD experiment. Figure \ref{fig-uniform-beam} shows the numerical result. The dispersion relation in wave number space is displayed in Figure \ref{fig-uniform-beam}a, color coded by the real wave frequency. It is seen that the real wave frequency near resonance has a small down-shift compared to that of the cold plasma dispersion relation. Away from the resonance, the beam has a negligible effect on the dispersion relation. The corresponding wave growth rate, i.e., the imaginary part of the frequency, is shown in wave number space in Figure \ref{fig-uniform-beam}b. Three resonance modes (i.e., cyclotron resonance, Landau resonance and anomalous cyclotron resonance) show up prominently. Each of the resonance modes has a wide distribution in wave normal angle. The cyclotron resonance mode is counter-streaming with the beam with $-2 < k_z c / \omega_{pe} < 0$. The Landau resonance mode and anomalous cyclotron mode are co-streaming with the beam. Note that the two branches in the region $0 < k_z c / \omega_{pe} < 2$ with appreciable growth rate belong to the Landau resonance. The anomalous cyclotron resonance mode is located in the region with a large parallel wave number $k_z c / \omega_{pe} > 2$.

\section{Matching the eigenmodes at the boundary}\label{sec:match_boundary_condition}
Here we match the eigenmodes inside and outside the beam at the boundary. The electric field on the side of the beam region with $x>a$ has the form
\begin{eqnarray}
\hat{\mathbf{E}}^{out} (x>a) = \begin{pmatrix}
\tilde{E}_x^{out} \\
\tilde{E}_y^{out} \\
\tilde{E}_z^{out}
\end{pmatrix} \exp(-i \omega t + i k_z z + i k_y y + i k_x^{out} x) ,
\end{eqnarray}
whereas the electric field on the other side of the beam with $x < -a$ has the form
\begin{eqnarray}
\hat{\mathbf{E}}^{out} (x<-a) = \begin{pmatrix}
-\tilde{E}_x^{out} \\
-\tilde{E}_y^{out} \\
 \tilde{E}_z^{out}
\end{pmatrix} \exp(-i \omega t + i k_z z + i k_y y - i k_x^{out} x) .
\end{eqnarray}
Here $k_y$ has a continuous spectrum and is set to $0$ hereafter for simplicity. The sign change for the $E_{x}$ and $E_y$ components is due to the fact that the orientation of $k_x$ is opposite in the region of $x>a$ compared to $x<-a$. Note that $k_x^{out} > 0$ indicates outgoing wave fronts away from the beam and $k_x^{out} < 0$ indicates incoming wave fronts toward the beam. The sign of $k_x^{out}$ should be chosen in a physically meaningful way such that the Poynting flux is directed away from the beam. The electric field inside the beam region has the form
\begin{eqnarray}
\hat{\mathbf{E}}^{in} = \left[ \begin{pmatrix}
\tilde{E}_x^{+} \\
\tilde{E}_y^{+} \\
\tilde{E}_z^{+}
\end{pmatrix} \exp(i k_x^{in} x) + \begin{pmatrix}
\tilde{E}_x^{-} \\
\tilde{E}_y^{-} \\
\tilde{E}_z^{-}
\end{pmatrix} \exp(-i k_x^{in} x) \right] \exp(-i \omega t + i k_z z) .
\end{eqnarray}
The boundary condition imposed is the continuity of tangential electric field across the boundary, i.e., $E_y^{in} |_{x = \pm a} = E_y^{out} |_{x = \pm a}$ and $E_z^{in} |_{x = \pm a} = E_z^{out} |_{x = \pm a}$. Note that the continuity of $B_x$ across the boundary is equivalent to the continuity of $E_y$ because of $B_x = -(k_z c / \omega) E_y$. This boundary condition can be written explicitly as
\begin{eqnarray}
\begin{cases}\label{ey_cont}
E_y^+ \exp(i k_x^{in} a) + E_y^- \exp(-i k_x^{in} a) = E_y^{out} \exp(i k_x^{out} a) \\
E_y^+ \exp(-i k_x^{in} a) + E_y^- \exp(i k_x^{in} a) = - E_y^{out} \exp(i k_x^{out} a)
\end{cases} 
\end{eqnarray}
and
\begin{eqnarray}
\begin{cases}\label{ez_cont}
E_z^+ \exp(i k_x^{in} a) + E_z^- \exp(-i k_x^{in} a) = E_z^{out} \exp(i k_x^{out} a) \\
E_z^+ \exp(-i k_x^{in} a) + E_z^- \exp(i k_x^{in} a) = E_z^{out} \exp(i k_x^{out} a)
\end{cases} .
\end{eqnarray}
Adding the conditions at $x=\pm a$ in equation \eqref{ey_cont} and subtracting the conditions at $x=\pm a$ in equation \eqref{ez_cont}, respectively, we obtain
\begin{eqnarray}
(E_y^+ + E_y^-) \cos(k_x^{in} a) = 0 , \\
(E_z^+ - E_z^-) \sin(k_x^{in} a) = 0 .
\end{eqnarray}
The only possible solution is
\begin{eqnarray}
E_y^+ &=& - E_y^- , \label{ey-odd} \\
E_z^+ &=& E_z^- .
\end{eqnarray}
This pair of equations means that the mode structure of $E_y$ is odd and that of $E_z$ is even. In fact, the symmetry also requires that the mode structure of $E_x$ is odd. Equations \eqref{ey_cont} and \eqref{ez_cont} can be rewritten to relate the field inside and outside the beam through
\begin{eqnarray}
2 i E_y^{+} \sin(k_x^{in} a) &=& E_y^{out} \exp(i k_x^{out} a) , \label{eyinout} \\
2 E_z^+ \cos(k_x^{in} a) &=& E_z^{out} \exp(i k_x^{out} a) . \label{ezinout}
\end{eqnarray}
In principle, one can write two of the electric field components in terms of the third component, e.g.,
\begin{eqnarray}
\frac{E_x^+}{E_y^+} &=& r^{+} , \\
\frac{E_z^+}{E_y^+} &=& s^{+} . \label{ezpdeyp}
\end{eqnarray}
The symmetry requires that
\begin{eqnarray}
\frac{E_x^-}{E_y^-} &=& r^{+} , \\
\frac{E_z^-}{E_y^-} &=& -s^{+} . \label{ezmddym}
\end{eqnarray}
Outside the beam, one has
\begin{eqnarray}
\frac{E_x^{out}}{E_y^{out}} &=& r^{out} , \\
\frac{E_z^{out}}{E_y^{out}} &=& s^{out} . \label{Eratio_out}
\end{eqnarray}
Using equations \eqref{ezpdeyp} and \eqref{Eratio_out}, we can write equations \eqref{eyinout} and \eqref{ezinout} as a linear system about $E_y^+$ and $E_y^{out}$. The determinant of this linear system must vanish, i.e,
\begin{eqnarray}\label{boundary_cond}
i s^{out} \tan(k_x^{in} a) = s^+ .
\end{eqnarray}
This is the final boundary condition obtained by matching the eigenmodes inside and outside the electron beam.

\begin{figure}[h]
\centering
\includegraphics[width=0.675\textwidth]{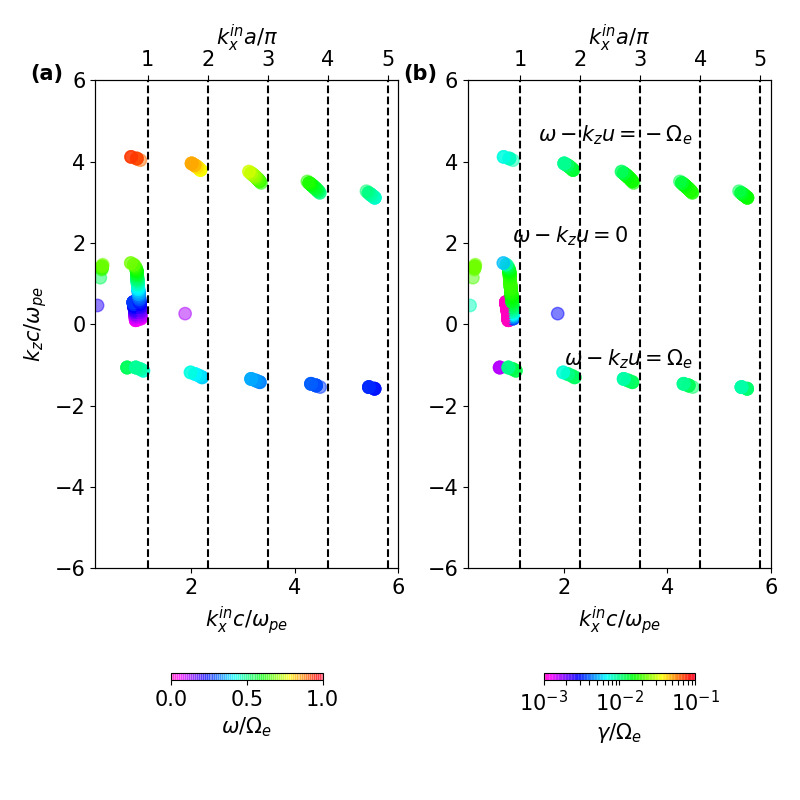}
\caption{(a) The solution of linearly unstable whistler eigenmodes for a finite electron beam in the wave number space. Each solution is color coded by the real wave frequency, with the color bar shown on the bottom. The vertical dashed blue lines indicate $k_x^{in} a = n \pi$ ($n$ is an integer) to highlight the quantization of $k_x^{in}$. (b) The corresponding growth rate of each wave mode.}
\label{fig-finite-beam}
\end{figure}

The dispersion relation inside the beam is
\begin{eqnarray}\label{disp_in}
\det(\mathbf{M}^{in}) = 0 ,
\end{eqnarray}
where
\begin{eqnarray}
\mathbf{M}^{in} = \frac{c^2}{\omega^2}\mathbf{k}^{in}\mathbf{k}^{in} - \frac{(k^{in})^2 c^2}{\omega^2} \mathbf{I}_3 + \boldsymbol{\epsilon}_c + \boldsymbol{\chi}_b
\end{eqnarray}
with $\mathbf{k}^{in} = (k_x^{in}, 0, k_z)$. The dispersion relation outside the beam is
\begin{eqnarray}\label{disp_out}
\det(\mathbf{M}^{out}) = 0 ,
\end{eqnarray}
where
\begin{eqnarray}
\mathbf{M}^{out} = \frac{c^2}{\omega^2}\mathbf{k}^{out}\mathbf{k}^{out} - \frac{(k^{out})^2 c^2}{\omega^2} \mathbf{I}_3 + \boldsymbol{\epsilon}_c
\end{eqnarray}
with $\mathbf{k}^{out} = (k_x^{out}, 0, k_z)$. Translational symmetry in the $z$ direction implies that the spectrum of $k_z$ is continuous. Therefore, for a given $k_z$, equations \eqref{boundary_cond}, \eqref{disp_in} and \eqref{disp_out} are solved simultaneously for values of $\omega$, $k_x^{in}$ and $k_x^{out}$ in the complex domain. To be able to make a comparison with the result of an infinite beam in section \ref{sec:linear_infinite_beam}, the parameters of the background plasma and the beam are kept the same. The half beam width is $10$ times the beam electron gyro-radius, i.e., $a = 10 v_{\perp 0}/ \Omega_e$, which is similar to that of our experiment. An overview of the numerical results is shown in Figure \ref{fig-finite-beam}. The solutions are plotted with respect to the parallel wave number $k_z$ and the real part of the perpendicular wave number inside the beam $k_x^{in}$, color coded by the real wave frequency in Figure \ref{fig-finite-beam}a and by the growth rate in Figure \ref{fig-finite-beam}b. It is seen that the perpendicular wave number inside the beam ($k_x^{in}$) is quantized, resulting from the constraint imposed by the boundary condition \eqref{boundary_cond}. The vertical dashed blue lines $k_x^{in} a = n \pi$ ($n$ is an integer) are plotted as a reference. The solutions with appreciable growth are clustered around three resonance lines $\omega - k_z u = n\Omega_e$ with $n=1, 0, -1$ from bottom to top. These results will be quantitatively compared to the experimental results in the next section.

\begin{figure}[h]
\centering
\includegraphics[width=0.675\textwidth]{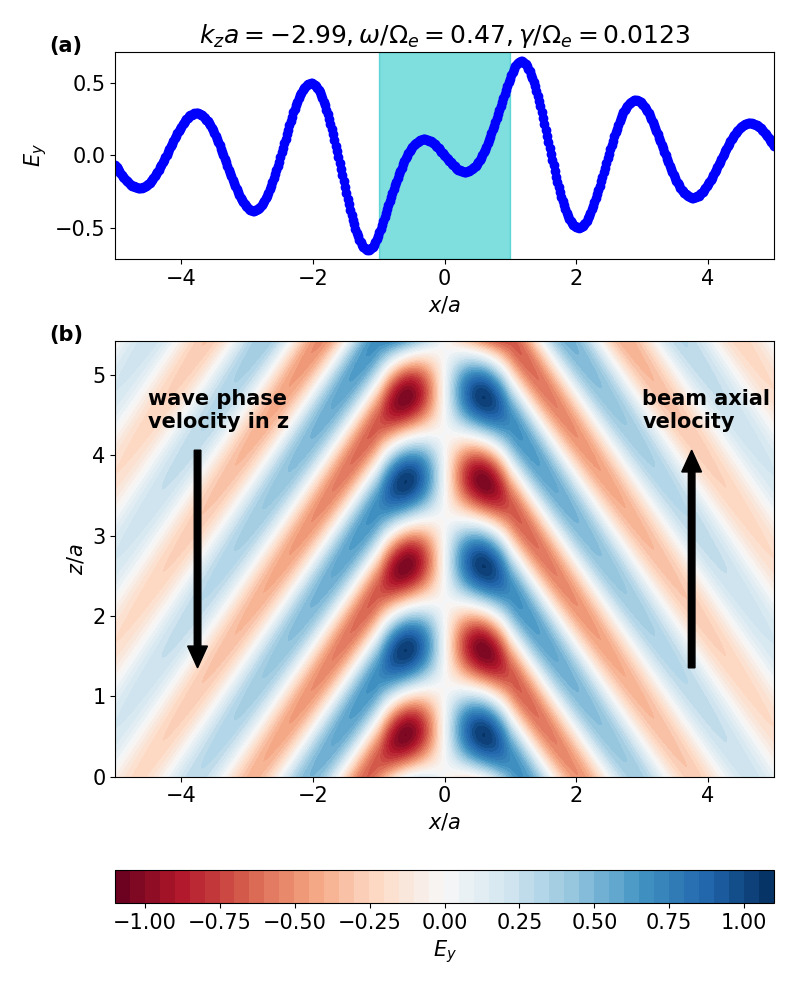}
\caption{A chosen mode structure excited by cyclotron resonance. The corresponding parallel wave number $k_z$, wave frequency $\omega$ and wave growth rate $\gamma$ is displayed on the top. (a) The $y$ component of the electric field $E_y$ as a function of position $x$. The shaded region indicates where the beam is located. (b) The mode structure of $E_y$ in the $x - z$ plane.}
\label{fig-mode-cyclotron}
\end{figure}

\begin{figure}[h]
\centering
\includegraphics[width=0.675\textwidth]{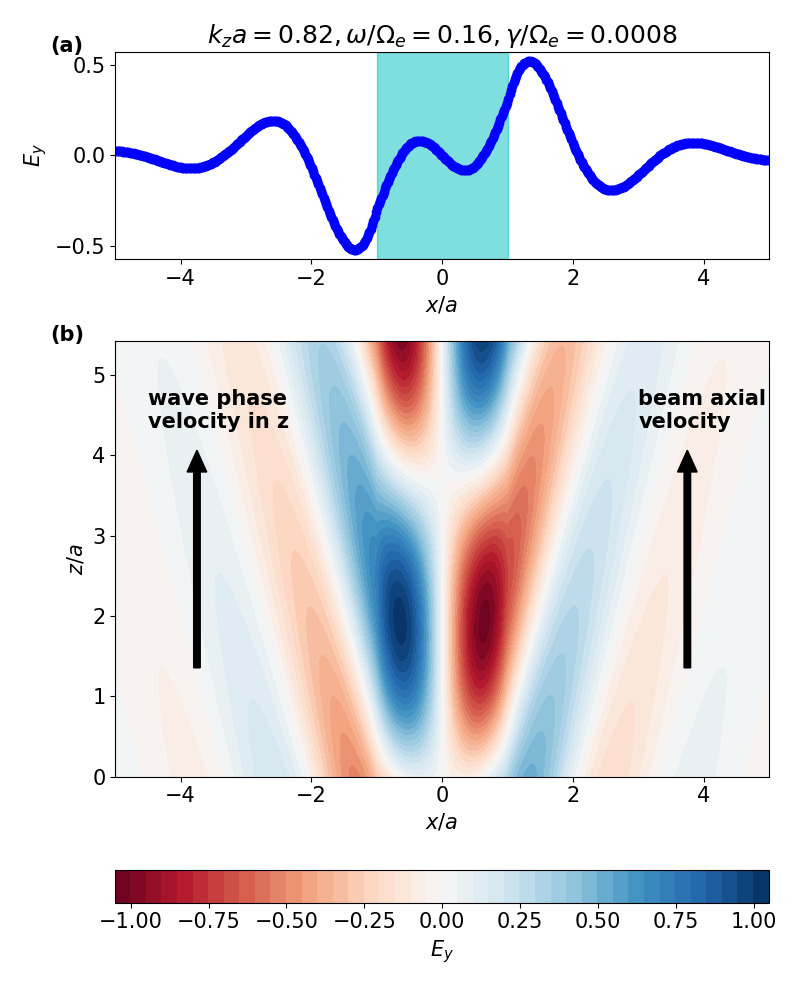}
\caption{A chosen mode structure excited by Landau resonance. The format is similar to that of Figure \ref{fig-mode-cyclotron}.}
\label{fig-mode-Landau}
\end{figure}

\begin{figure}[h]
\centering
\includegraphics[width=0.675\textwidth]{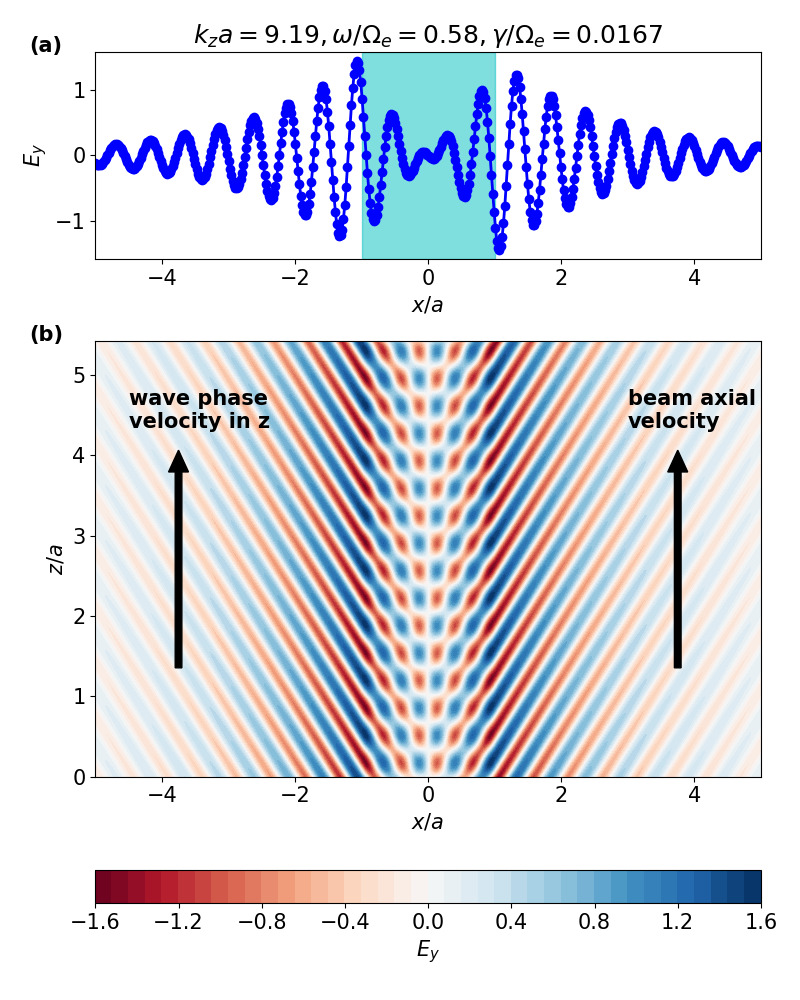}
\caption{A chosen mode structure excited by anomalous cyclotron resonance. The format is similar to that of Figure \ref{fig-mode-cyclotron}.}
\label{fig-mode-anomalous}
\end{figure}

Three representative mode structures of $E_y$, taken from the cyclotron resonance mode, Landau resonance mode and anomalous resonance mode are shown in Figures \ref{fig-mode-cyclotron}, \ref{fig-mode-Landau} and \ref{fig-mode-anomalous} respectively. In all three plots, $E_y$ with respect to $x$ is displayed in the upper panel and an image of $E_y$ in the $x - z$ plane is shown in the lower panel. As shown in treating the boundary condition, $E_y$ is an odd mode. The amplitude of $E_y$ peaks in the vicinity of the boundary and decays to zero as $x \to \pm \infty$. The beam width is comparable to the perpendicular wave length inside the beam. It is worthy to note that the wave front outside the beam is oblique and converges toward the beam while the Poynting flux flows out of the beam. In terms of wave propagation, the cyclotron resonance mode is counter-streaming with the beam (Figure \ref{fig-mode-cyclotron}), whereas both the Landau resonance mode and the anomalous cyclotron resonance mode are co-streaming with the electron beam (Figures \ref{fig-mode-Landau} and \ref{fig-mode-anomalous}). These plots show the benefit of having a radially extended beam, as the waves are allowed to grow to larger amplitudes before exiting the beam region.

\section{Comparison with the experiment}\label{sec:experiment}

\begin{figure}[h]
\centering
\includegraphics[width=0.675\textwidth]{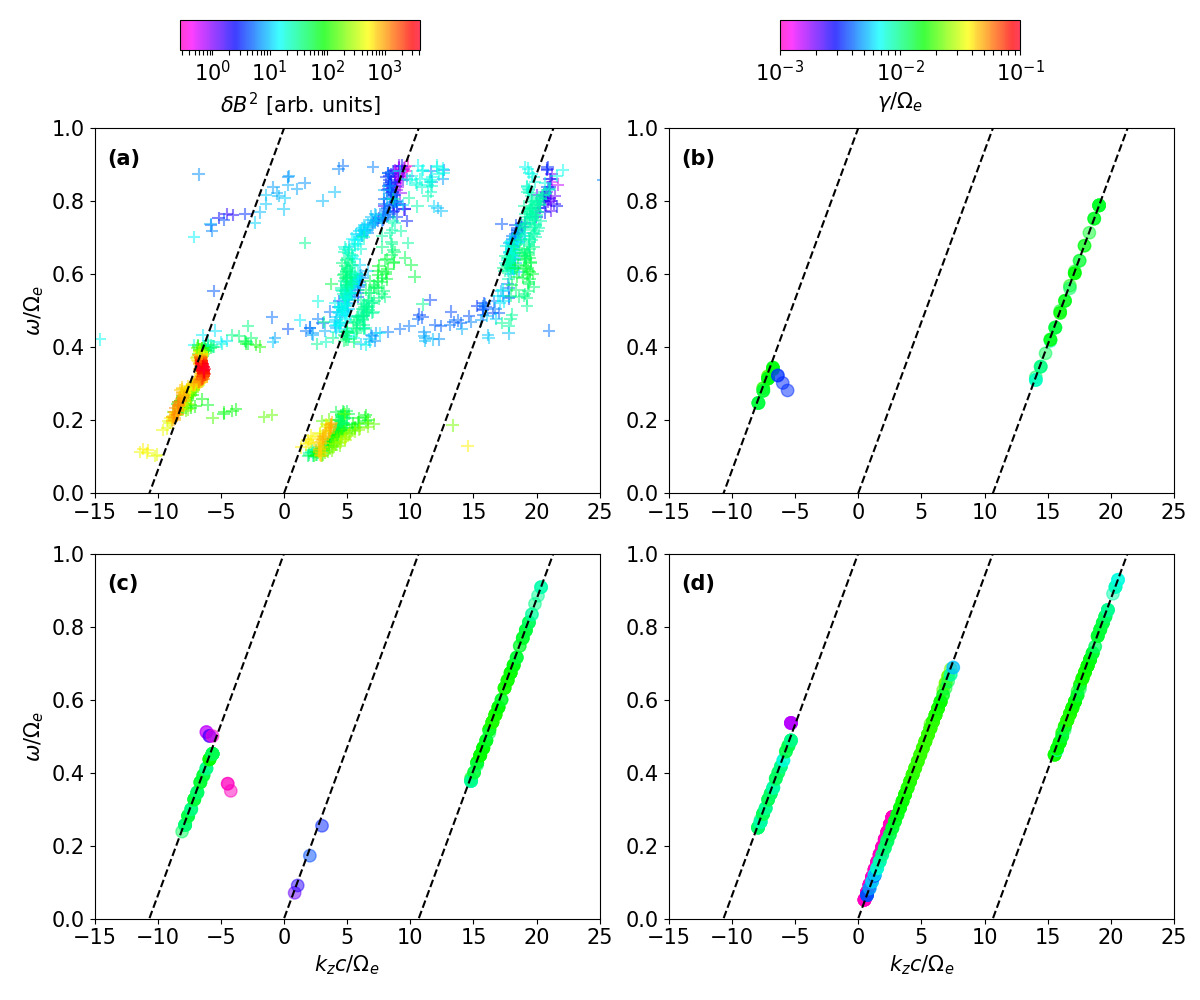}
\caption{The comparison between the experimental result and the linear instability analysis. (a) The experimentally measured parallel wave number $k_z$ and wave frequency $\omega$. Each data point is color coded by the power spectral density. (b), (c), (d) The theoretical solution as a function of the parallel wave number and the wave frequency, color coded by the growth rate. In the experiment, $\omega_{pe} / \Omega_{e}$ was measured to be $9.6$, while in the linear instability analysis, $\omega_{pe} / \Omega_{e}$ is set as $9.6$, $6.0$ and $5.0$ for (b), (c) and (d), respectively. Three dashed black lines in each panel, from left to right, represents $\omega - k_z u = n\Omega_e$ with $n=1, 0, -1$. Panel (d) is a representation of the nominal case in Figure \ref{fig-finite-beam}.}
\label{fig-resonances}
\end{figure}

To interpret our experimental results with the linear instability analysis, we organize the experimental results (see Ref.~\onlinecite{an2016resonant} for details) into the plane of  $k_z - \omega$ as shown in Figure \ref{fig-resonances}(a). All the parameters in the experiment are the same as that in Figure \ref{fig-finite-beam} except that $\omega_{pe} / \Omega_{e} = 9.6$. The linear instability analysis is shown in Figures \ref{fig-resonances}(b), \ref{fig-resonances}(c) and \ref{fig-resonances}(d), corresponding to $\omega_{pe} / \Omega_{e} = 9.6, 6.0, 5.0$, respectively. Note that the linear instability analysis in Figure \ref{fig-resonances}(b) has the same parameters as that in the experiment, whereas the analysis in Figures \ref{fig-resonances}(c) and \ref{fig-resonances}(d) varies $\omega_{pe} / \Omega_e$ and keeps other parameters unchanged. In the experiment, whistler waves are excited primarily due to three basic resonance regimes simultaneously: the normal cyclotron resonance mode ($0.2 < \omega / \Omega_e < 0.4$), Landau resonance ($0.1 < \omega / \Omega_e < 0.25$ and $0.4 < \omega / \Omega_e < 0.9$) and first-order anomalous cyclotron resonance mode ($0.4 < \omega / \Omega_e < 0.9$). The location of both cyclotron resonance and anomalous cyclotron resonance in $k_z$-$\omega$ space is accurately captured by the linear analysis in Figure \ref{fig-resonances}(b). But Landau resonance is not well reproduced. As $\omega_{pe} / \Omega_e$ decreases, Landau resonance starts to emerge as seen in Figures \ref{fig-resonances}(c) and \ref{fig-resonances}(d). It can be shown that \cite{starodubtsev1999resonantc, an2017electrostatic} Landau resonance between whistler waves and beam electrons only occurs below some critical plasma density. From the refractive index surface of whistler waves, there exists a minimum $k_z$ for a given frequency 
\begin{eqnarray}
k_z^{\min} = \begin{cases}
\dfrac{\omega_{pe}}{c}\dfrac{2\omega}{\Omega_e} & \omega < \dfrac{\Omega_e}{2} \\
\dfrac{\omega_{pe}}{c} \sqrt{\dfrac{\omega}{\Omega_e-\omega}} & \omega > \dfrac{\Omega_e}{2}
\end{cases}. 
\end{eqnarray}  
In order to have Landau resonance between beam electrons and whistler waves, the resonant wave number must exceed $k_z^{\min}$. That is $$\dfrac{\omega}{u} > k_z^{\min}. $$ There exists a critical value of $\omega_{pe}/\Omega_e$, above which Landau resonance does not occur. This critical value is 
\begin{eqnarray}
\left(\dfrac{\omega_{pe}}{\Omega_e}\right)_{\mbox{critical}} = \begin{cases}
\dfrac{c}{2u} & \omega < \dfrac{\Omega_e}{2} \\
\sqrt{\dfrac{\omega}{\Omega_e} \left(1-\dfrac{\omega}{\Omega_e}\right)}\dfrac{c}{u} & \omega > \dfrac{\Omega_e}{2}
\end{cases}. 
\end{eqnarray}  
For typical parameters in the experiment, i.e., $u/c=0.1$ and $\omega/\Omega_e=0.5$, the critical value of $\omega_{pe}/\Omega_e$ is $5$. For
Landau resonance in the high density regime $\omega_{pe} / \Omega_e = 9.6$ in the experiment, we do not have a satisfactory answer in the current stage.

\section{Summary and Discussion}\label{sec:sum-discuss}
The linear unstable whistler wave eigenmodes excited by a gyrating electron beam in a slab geometry are studied here. A linear instability analysis is first performed to find the resonance modes and the associated wave growth rate for an infinite beam superposed on a cold background plasma. By matching the eigenmodes of whistler waves at the boundaries for a finite beam, the system of equation become closed and a complex wave frequency can be found for each wave mode and the corresponding mode structure can be constructed. It is shown that the perpendicular wave number inside the beam with an appreciable growth is quantized analogously to a waveguide. The symmetry of $x$ and $y$ components of the wave electric field is odd in the transverse direction whereas that of $z$ component of the wave electric field is even in the transverse. The wave electric field peaks in the vicinity of the boundary and can leak out of the electron beam, decaying to zero at infinity.

It should be pointed out that the calculation in this study is only applicable to a tenuous beam, i.e., $n_b \ll n_0$. The ratio $n_b / n_0$ is the range from $0.5 \times 10^{-3}$ to $3.5 \times 10^{-3}$ in the experiment \cite{an2016resonant} such that this condition is satisfied. For one thing, the validity of linear kinetic theory relies on the wave growth rate being much smaller than the wave frequency so that only first order terms are retained, which requires $n_b \ll n_0$. In addition, the initial setup of the system is not in an equilibrium state. The continuity of the electric displacement at the boundary is ensured only when $n_b \ll n_0$. This means that the beam does not change the dispersion properties of the background plasma significantly but only provides a small correction to the real frequency as well as a wave growth rate (complex frequency). The enforcement of the continuity of the electric displacement into the computation leads to an over-determined system\footnote{For a given parallel wave number, this system has four equations but only three unknowns. Four equations are the dispersion relations inside and outside the beam, one boundary condition from the continuity of tangential electric field and the other boundary condition from the continuity of the electric displacement. The three unknowns are the wave frequency, the perpendicular wave numbers inside and outside the beam.}. The problem setup is self-consistent only when the beam is tenuous. Such a dilemma can be reconciled if we consider a finite beam with a smooth density profile. Under the setup of a smooth beam, the only boundary condition is that all the wave fields vanish at infinity. By expanding the wave fields and the electron distribution around the gyro-orbit of beam electrons, the linearized Vlasov equation can be integrated properly. A set of differential equations can be obtained, which defines the eigenvalue problem. This type of geometry, which is both more realistic and more complex than the one considered in this manuscript, will be studied in the future.

\section{Supplementary Material}
See supplementary material for the detailed calculation for the susceptibility tensor of a gyrating electron beam.


%
%

%

\begin{acknowledgments}
This work was funded by the NASA Grant No.~NNX16AG21G. We would like to acknowledge high-performance computing support from Cheyenne (doi:10.5065/D6RX99HX) provided by NCAR's Computational and Information Systems Laboratory, sponsored by the National Science Foundation. We wish to thank Prof.~George J.~Morales for the suggestion of performing such a study. X.~A.~would like to thank Dr.~Anton V.~Artemyev for helpful discussions.
\end{acknowledgments}

%

\pagebreak
\widetext
\begin{center}
\textbf{\large Supplementary Material to ``Linear unstable whistler eigenmodes excited by a finite electron beam''}
\end{center}
\setcounter{equation}{0}
\setcounter{figure}{0}
\setcounter{table}{0}
\setcounter{page}{1}
\makeatletter
\renewcommand{\theequation}{S\arabic{equation}}
\renewcommand{\thefigure}{S\arabic{figure}}

Here the susceptibility tensor for a gyrating electron beam is calculated analytically. To begin with, the linearized Vlasov equation is
\begin{eqnarray} \label{linear_vlasov}
\frac{d \hat{f}_b}{d t} -\frac{e}{m} \left( \hat{\mathbf{E}} + \frac{\mathbf{v}\times \hat{\mathbf{B}}}{c} \right) \cdot \frac{\partial f_{0b}}{\partial \mathbf{v}} = 0 ,
\end{eqnarray}
where $\hat{f}_b$ is the perturbed distribution function and $\hat{\mathbf{E}}$, $\hat{\mathbf{B}}$ are the perturbed fields. The operator $d / dt$ denotes
\begin{eqnarray}
\frac{d}{dt} = \frac{\partial}{\partial t} + \mathbf{v} \cdot \boldsymbol{\nabla} -\frac{e}{m} \left( \frac{\mathbf{v}\times \mathbf{B}_0}{c} \right) \cdot \frac{\partial}{\partial \mathbf{v}} ,
\end{eqnarray}
where $\mathbf{B}_0$ is the background magnetic field. We consider perturbations of the form
\begin{eqnarray}
\hat{\mathbf{E}} &=& \tilde{\mathbf{E}} \exp(-i\omega t + i k_x x + i k_y y + i k_z z) , \\
\hat{\mathbf{B}} &=& \tilde{\mathbf{B}} \exp(-i\omega t + i k_x x + i k_y y + i k_z z) .
\end{eqnarray}
Equation \eqref{linear_vlasov} can be integrated along its characteristics, i.e., the unperturbed helical orbits of electrons. This integral can be calculated as
\begin{eqnarray} \label{deltaf}
\hat{f}_b &=& \frac{e}{m} \int_{-\infty}^{t} d t^{\prime} \exp(-i\omega t^\prime + i k_x x^\prime + i k_y y^\prime + i k_z z^\prime) \tilde{S} .
\end{eqnarray}
Here the integral kernel $\tilde{S}$ is
\begin{eqnarray} \label{int_kernel}
\tilde{S} &=& \left( \tilde{\mathbf{E}} + \frac{\mathbf{v}^\prime \times \tilde{\mathbf{B}}}{c} \right) \cdot \frac{\partial f_{0b}(\mathbf{v}^\prime)}{\partial \mathbf{v}^\prime} \nonumber \\
&=& \left( v_x^\prime \tilde{E}_x + v_y^\prime \tilde{E}_y \right) f_{0b\perp} + \tilde{E}_z f_{0bz} \nonumber \\
& & + \left[ \frac{v_x^\prime}{c} (v_y^\prime \tilde{B}_z - v_z^\prime \tilde{B}_y) + \frac{v_y^\prime}{c} (v_z^\prime \tilde{B}_x - v_x^\prime \tilde{B}_z)  \right] f_{0b\perp} + \frac{1}{c} (v_x^\prime \tilde{B}_y - v_y^\prime \tilde{B}_x) f_{0bz} \nonumber \\
&=& \tilde{E}_x v_x^\prime f_{0b\perp} + \tilde{E}_y v_y^\prime f_{0b\perp} + \tilde{E}_z f_{0bz} \nonumber \\
& & + \tilde{B}_x \left[ \frac{v_y^\prime}{c} (v_z f_{0b\perp} - f_{0bz}) \right] + \tilde{B}_y \left[ -\frac{v_x^\prime}{c} (v_z f_{0b\perp} - f_{0bz}) \right] .
\end{eqnarray}
We define
\begin{eqnarray}
f_{0b\perp} = \frac{1}{v_\perp} \frac{\partial f_{0b}}{\partial v_\perp} ,\\
f_{0bz} = \frac{\partial f_{0b}}{\partial v_z} .
\end{eqnarray}
Note that $v_{\perp}$ and $v_{z}$ are two constants along the particle orbit. Using Faraday's equation, we can write $\tilde{B}_x$ and $\tilde{B}_y$ as
\begin{eqnarray}
\tilde{B}_x &=& \frac{k_y c}{\omega} \tilde{E}_z - \frac{k_z c}{\omega} \tilde{E}_y , \\
\tilde{B}_y &=& \frac{k_z c}{\omega} \tilde{E}_x - \frac{k_x c}{\omega} \tilde{E}_z .
\end{eqnarray}
Replacing $\tilde{B}_x$ and $\tilde{B}_y$ with the electric field, one obtains
\begin{eqnarray}
\begin{split}
\tilde{S} =& \tilde{E}_x v_x^\prime \left[ f_{0b\perp} -\frac{k_z}{\omega} (v_z f_{0b\perp} - f_{0bz}) \right] \\
 +& \tilde{E}_y v_y^\prime \left[ f_{0b\perp} - \frac{k_z}{\omega} (v_z f_{0b\perp} - f_{0bz}) \right] \\
 +& \tilde{E}_z \left[ f_{0bz} + \frac{k_y v_y^\prime}{\omega} (v_z f_{0b\perp} - f_{0bz}) + \frac{k_x v_x^\prime}{\omega} (v_z f_{0b\perp} - f_{0bz}) \right] .
\end{split} 
\end{eqnarray}
The unperturbed particle orbit can be solved as 
\begin{eqnarray}
v_x^\prime &=& v_x \cos(\Omega_e \tau) + v_y \sin(\Omega_e \tau) , \\
v_y^\prime &=& -v_x \sin(\Omega_e \tau) + v_y \cos(\Omega_e \tau) , \\
v_z^\prime &=& v_z , \\
x^\prime &=& -\frac{v_x}{\Omega_e} \sin(\Omega_e \tau) - \frac{v_y}{\Omega_e} [1 - \cos(\Omega_e \tau)] + x , \\
y^\prime &=& \frac{v_x}{\Omega_e} [1 - \cos(\Omega_e \tau)] - \frac{v_y}{\Omega_e} \sin(\Omega_e \tau) + y , \\
z^\prime &=& -v_z \tau + z .
\end{eqnarray}
where $\tau = t - t^\prime$. Note that $\Omega_e = e B_0/ m c$ is the unsigned electron cyclotron frequency. The particle orbit reaches $\mathbf{v}^\prime = \mathbf{v}$ and $\mathbf{x}^\prime = \mathbf{x}$ at $t^\prime = t$. Using the particle orbit, we further write $\tilde{S}$ as
\begin{eqnarray}
\begin{split}
\tilde{S}(\tau) =& \tilde{E}_x(x) \left( v_x \cos(\Omega_e \tau) + v_y \sin(\Omega_e \tau) \right) \left[ f_{0b\perp} -\frac{k_z}{\omega} (v_z f_{0b\perp} - f_{0bz}) \right] \\
+& \tilde{E}_y(x) \left( -v_x \sin(\Omega_e \tau) + v_y \cos(\Omega_e \tau) \right) \left[ f_{0b\perp} -\frac{k_z}{\omega} (v_z f_{0b\perp} - f_{0bz}) \right] \\
+& \tilde{E}_z(x) \left[ f_{0bz} + \left( -v_x \sin(\Omega_e \tau) + v_y \cos(\Omega_e \tau) \right) \frac{k_y}{\omega} (v_z f_{0b\perp} - f_{0bz}) \right. \\
+& \left. \left( v_x \cos(\Omega_e \tau) + v_y \sin(\Omega_e \tau) \right) \frac{k_x}{\omega} (v_z f_{0b\perp} - f_{0bz}) \right] .
\end{split} 
\end{eqnarray}
To write $\tilde{S}$ in a more compact form, we define
\begin{eqnarray}
v_x &=& v_\perp \cos \phi ,\\
v_y &=& v_\perp \sin \phi ,\\
k_x &=& k_\perp \cos \psi ,\\
k_y &=& k_\perp \sin \psi ,\\
g_\perp &=& \frac{\partial f_{0b}}{\partial v_\perp} = v_\perp f_{0b\perp} ,\\
g_z &=& \frac{\partial f_{0b}}{\partial v_z} = f_{0bz} .
\end{eqnarray}
$\tilde{S}$ becomes
\begin{eqnarray}
\begin{split}
\tilde{S}(\tau) =& \tilde{E}_x(x) \cos(\phi - \Omega_e \tau) \left[ g_{\perp} + \frac{k_z}{\omega}(v_\perp g_{z} - v_z g_{\perp}) \right] \\
+& \tilde{E}_y(x) \sin(\phi - \Omega_e \tau) \left[ g_{\perp} + \frac{k_z}{\omega}(v_\perp g_{z} - v_z g_{\perp}) \right] \\
+& \tilde{E}_z(x) \left[ g_{z} - \cos(\phi - \psi - \Omega_e \tau) \frac{k_\perp}{\omega} (v_{\perp} g_{z} - v_z g_{\perp}) \right] .
\end{split}
\end{eqnarray}
Now we express the phase factor in equation \eqref{deltaf} as a function of $\tau$
\begin{equation}
\begin{split}
-i\omega t^\prime + i k_x x^\prime + i k_y y^\prime + i k_z z^\prime =& -i\omega (t - \tau) + i k_x \left[ x + \frac{v_\perp}{\Omega_e} \left( -\sin\phi + \sin(\phi - \Omega_e \tau) \right) \right] \\
&+ i k_y \left[y + \frac{v_\perp}{\Omega_e}\left(\cos\phi - \cos(\phi - \Omega_e \tau) \right) \right] + i k_z (z - v_z \tau) \\
=& -i \omega t + i k_x x + i k_y y + i k_z z + i(\omega - k_z v_z) \tau \\
&+ i \frac{k_x v_\perp}{\Omega_e} \left( -\sin\phi + \sin(\phi - \Omega_e \tau) \right) + i \frac{k_y v_\perp}{\Omega_e}\left(\cos\phi - \cos(\phi - \Omega_e \tau) \right) \\
=& -i \omega t + i k_x x + i k_y y + i k_z z + i(\omega - k_z v_z) \tau \\
&+ i \frac{k_\perp v_\perp}{\Omega_e} \left( -\sin(\phi - \psi) + \sin(\phi - \psi - \Omega_e \tau) \right) .
\end{split} 
\end{equation}
Therefore the perturbed distribution can be integrated over $\tau$ as
\begin{equation}\label{fb_int_tau}
\begin{split}
\tilde{f}_b =& \frac{e}{m}  \exp(-i \omega t + i k_x x + i k_y y + i k_z z)  \\
&\times \int_0^\infty d \tau \exp\left[i(\omega - k_z v_z) \tau + i \frac{k_\perp v_\perp}{\Omega_e} \left( -\sin(\phi - \psi) + \sin(\phi - \psi - \Omega_e \tau) \right) \right] \tilde{S}(\tau) .
\end{split} 
\end{equation}
Using the Jacobi-Anger expansion $e^{i z \sin \psi} = \sum\limits_{m=-\infty}^{\infty} J_m(z) e^{i m \psi}$, we have
\begin{equation}
\exp\left(- i \frac{k_\perp v_\perp}{\Omega_e} \sin(\phi-\psi) \right) = \sum\limits_{m=-\infty}^{\infty} (-1)^m J_m\left(\frac{k_\perp v_\perp}{\Omega_e}\right) \exp(i m (\phi-\psi)) ,
\end{equation}
\begin{equation}
\exp\left( i \frac{k_\perp v_\perp}{\Omega_e} \sin(\phi - \psi - \Omega_e \tau)\right) = \sum\limits_{n=-\infty}^{\infty} J_n\left(\frac{k_\perp v_\perp}{\Omega_e}\right) \exp(i n (\phi - \psi - \Omega_e \tau)) .
\end{equation}
Hereafter $k_y = 0$ or $\psi = 0$ is assumed using the symmetry perpendicular to the background magnetic field. Another simplification is averaging $\hat{f}_b$ and its velocity moments over $\phi$ in velocity space when calculating the charge density and current density. The following identities are useful in this procedure
\begin{equation}
\begin{split}
& \frac{1}{2\pi} \int_0^{2\pi} d \phi \exp\left[i \frac{k_\perp v_\perp}{\Omega_e}\left(-\sin\phi + \sin(\phi - \Omega_e \tau) \right) \right] \times \begin{pmatrix}
1 \\
\cos \phi \\
\sin \phi \\
\cos(\phi - \Omega_e \tau) \\
\sin(\phi - \Omega_e \tau) \\
\cos\phi \cos(\phi - \Omega_e \tau) \\
\cos\phi \sin(\phi - \Omega_e \tau) \\
\sin\phi \cos(\phi - \Omega_e \tau) \\
\sin\phi \sin(\phi - \Omega_e \tau)
\end{pmatrix} \\
&= \sum\limits_{n=-\infty}^{\infty}\exp(-i n \Omega_e \tau) \times \begin{pmatrix}
J_n^2 \\
\frac{n \Omega_e}{k_\perp v_\perp} J_n^2 \\
i J_n J_n^\prime \\
\frac{n \Omega_e}{k_\perp v_\perp} J_n^2 \\
-i J_n J_n^\prime \\
\frac{n^2 \Omega_e^2}{k_\perp^2 v_\perp^2} J_n^2 \\
-i \frac{n \Omega_e}{k_\perp v_\perp} J_n J_n^\prime \\
i \frac{n \Omega_e}{k_\perp v_\perp} J_n J_n^\prime \\
(J_n^\prime)^2
\end{pmatrix} .
\end{split}
\end{equation}
Here the argument of both Bessel function and its derivative is $\frac{k_\perp v_\perp}{\Omega_e}$. The integration over $\tau$ in equation \eqref{fb_int_tau} can be performed as
\begin{eqnarray}
\int_0^\infty d\tau \exp\left[i(\omega - k_z v_z - n \Omega_e) \tau \right] = \frac{i}{\omega - k_z v_z - n \Omega_e} ,
\end{eqnarray}
given that $\Im(\omega) > 0$. We denote the averaging procedure by $\langle \ \rangle_\phi$. The zeroth- and first-order moment of $\hat{f}_b$ becomes
\begin{eqnarray}\label{av_fb}
\begin{split}
\langle \hat{f}_b \rangle_\phi =& \frac{e}{m} \exp(-i \omega t + i k_x x + i k_z z) \sum\limits_{n=-\infty}^{\infty} \frac{1}{\omega - k_z v_z - n \Omega_e}  \\
&\times \left\lbrace \tilde{E}_x(x) \left( \frac{n \Omega_e}{k_\perp v_\perp} i J_n^2 \right) \left[g_{\perp} + \frac{k_z}{\omega}(v_\perp g_{z} - v_z g_{\perp}) \right] \right. \\
&+ \left. \tilde{E}_y(x) \left( J_{n} J_n^\prime \right) \left[g_{\perp} + \frac{k_z}{\omega}(v_\perp g_{z} - v_z g_{\perp}) \right] \right. \\
&+ \left. \tilde{E}_z(x) \left( i J_n^2 \right) \left[ \left( 1 - \frac{n \Omega_e}{\omega} \right) g_{z} + \left( \frac{n \Omega_e v_z}{\omega v_\perp} \right) g_\perp \right] \right\rbrace ,
\end{split}
\end{eqnarray}
\begin{eqnarray}\label{av_fb_vx}
\begin{split}
\langle v_x \hat{f}_b  \rangle_\phi =& \frac{e v_\perp}{m} \exp(-i \omega t + i k_x x + i k_z z) \sum\limits_{n=-\infty}^{\infty} \frac{1}{\omega - k_z v_z - n \Omega_e} \\
&\times \left\lbrace \tilde{E}_x(x) \left( \frac{n^2 \Omega_e^2}{k_\perp^2 v_\perp^2} i J_n^2 \right) \left[g_{\perp} + \frac{k_z}{\omega}(v_\perp g_{z} - v_z g_{\perp}) \right] \right. \\
&+ \left. \tilde{E}_y(x)  \left( \frac{n \Omega_e}{k_\perp v_\perp} J_n J_n^\prime \right) \left[g_{\perp} + \frac{k_z}{\omega}(v_\perp g_{z} - v_z g_{\perp}) \right] \right. \\
&+ \left. \tilde{E}_z(x)  \left( \frac{n \Omega_e}{k_\perp v_\perp} i J_n^2 \right) \left[ \left( 1 - \frac{n \Omega_e}{\omega} \right) g_{z} + \left( \frac{n \Omega_e v_z}{\omega v_\perp} \right) g_\perp \right] \right\rbrace ,
\end{split}
\end{eqnarray}
\begin{eqnarray}\label{av_fb_vy}
\begin{split}
\langle v_y \hat{f}_b  \rangle_\phi =& \frac{e v_\perp}{m} \exp(-i \omega t + i k_x x + i k_z z) \sum\limits_{n=-\infty}^{\infty} \frac{1}{\omega - k_z v_z - n \Omega_e} \\
&\times \left\lbrace \tilde{E}_x(x) \left( - \frac{n \Omega_e}{k_\perp v_\perp} J_n J_n^\prime \right) \left[g_{\perp} + \frac{k_z}{\omega}(v_\perp g_{z} - v_z g_{\perp}) \right] \right. \\
&+ \left. \tilde{E}_y(x) \left( i (J_n^\prime)^2 \right) \left[g_{\perp} + \frac{k_z}{\omega}(v_\perp g_{z} - v_z g_{\perp}) \right] \right. \\
&+ \left. \tilde{E}_z(x) \left( - J_n J_n^\prime \right) \left[ \left( 1 - \frac{n \Omega_e}{\omega} \right) g_{z} + \left( \frac{n \Omega_e v_z}{\omega v_\perp} \right) g_\perp \right] \right\rbrace ,
\end{split}
\end{eqnarray}
Knowledge of velocity moments now leads to the calculation of first-order plasma currents generated by the perturbed electron beam distribution, i.e.,
\begin{eqnarray}\label{jb}
\hat{\mathbf{j}}_b = -e \int_0^\infty 2\pi v_\perp d v_\perp \int_{-\infty}^{\infty} d v_z \langle \mathbf{v} \hat{f}_b \rangle_\phi .
\end{eqnarray}
Note that $\hat{\mathbf{j}}_b$ has the form $\hat{\mathbf{j}}_b = \tilde{\mathbf{j}}_b \exp(- i \omega t + i k_x x + i k_z z)$. From equation \eqref{av_fb}-\eqref{jb}, one observes that $\tilde{\mathbf{j}}_b$ can be expressed as a linear superposition of $\tilde{E}_x(x)$, $\tilde{E}_y(x)$ and $\tilde{E}_z(x)$, with the coefficients being integrals of gradients over velocity space, namely
\begin{eqnarray}\label{jb_lin_supp}
\begin{split}
\frac{4 \pi i}{\omega} \tilde{\mathbf{j}}_{b} = \boldsymbol{\chi}_b \cdot \tilde{\mathbf{E}} ,
\end{split}
\end{eqnarray}
where
\begin{eqnarray}\label{beam_susceptibility}
\begin{split}
\boldsymbol{\chi}_b =& \begin{pmatrix}
\chi_{xx}  & \chi_{xy}  & \chi_{xz} \\
\chi_{yx}  & \chi_{yy}  & \chi_{yz} \\
\chi_{zx}  & \chi_{zy}  & \chi_{zz}
\end{pmatrix} \\
=& \frac{\omega_{pb}^2}{\omega} \sum\limits_{n=-\infty}^{\infty} \int_0^\infty 2\pi v_\perp d v_\perp \int_{-\infty}^{\infty} d v_z \frac{1}{\omega - k_z v_z - n \Omega_e} \\
\times & \begin{pmatrix}
\frac{n^2 \Omega_e^2}{k_\perp^2 v_\perp^2} J_n^2 v_\perp U                                & -i \frac{n \Omega_e}{k_\perp v_\perp} J_n J_n^\prime v_\perp U  &   \frac{n \Omega_e}{k_\perp v_\perp} J_n^2 v_\perp W \\
i \frac{n \Omega_e}{k_\perp v_\perp} J_n J_n^\prime v_\perp U &  (J_n^\prime)^2 v_\perp U & i J_n J_n^\prime v_\perp W \\
\frac{n \Omega_e}{k_\perp v_\perp} J_n^2 v_z U                                    &  -i J_{n} J_n^\prime v_z U             &  J_n^2 v_z W
\end{pmatrix} .
\end{split}
\end{eqnarray}
We define
\begin{eqnarray}
U &=& g_{\perp} + \frac{k_z}{\omega}(v_\perp g_{z} - v_z g_{\perp}) , \\
W &=& \left( 1 - \frac{n \Omega_e}{\omega} \right) g_z + \frac{n \Omega_e v_z}{\omega v_\perp} g_\perp ,
\end{eqnarray}
where we have re-scale $g_\perp$ and $g_z$ to be $g_\perp = \frac{1}{n_b} \frac{\partial f_{0b}}{\partial v_\perp}$ and $g_z = \frac{1}{n_b} \frac{\partial f_{0b}}{\partial v_z}$, respectively. $n_b$ is beam density and $\omega_{pb}^2 = \frac{4 \pi n_b e^2}{m}$. Now we can calculate the specific susceptibility tensor for a beam ring distribution. We write the beam electron distribution as the following
\begin{eqnarray}
\begin{split}
f_{0b}(v_z, v_\perp) &= n_b P(v_z) Q(v_\perp) , \\
P(v_z) &= \delta(v_z - u) , \\
Q(v_\perp) &= \frac{1}{2 \pi v_\perp} \delta(v_\perp - v_{\perp 0}) .
\end{split}
\end{eqnarray}
Thus $U$ and $W$ can be rewritten as
\begin{eqnarray}
U &=& \left( 1 - \frac{k_z v_z}{\omega} \right) P Q_\perp + \frac{k_z v_\perp}{\omega} P_z Q , \\
W &=& \frac{n \Omega_e v_z}{\omega v_\perp} P Q_\perp + \left( 1 - \frac{n \Omega_e}{\omega} \right) P_z Q ,
\end{eqnarray}
where $Q_\perp = \frac{d Q}{d v_\perp}$ and $P_z = \frac{d P}{d v_z}$. To calculate the susceptibility tensor $\boldsymbol{\chi}_b$, two types of integration by parts are useful as the following
\begin{eqnarray}\label{int_id1}
\begin{split}
& \int_0^{\infty} d v_\perp \int_{-\infty}^{\infty} d v_z h(v_\perp) l(v_z) P Q_\perp \\
=& l(u) \int_0^{\infty} d v_\perp h(v_\perp) Q_\perp \\
=& l(u) [h(v_\perp) Q(v_\perp)]_{0}^{\infty} - l(u) \int_{0}^{\infty} d v_\perp Q \frac{d h}{d v_\perp} \\
=& -l(u) \frac{1}{2 \pi v_{\perp 0}} \frac{d h}{d v_\perp}(v_{\perp 0}) ,
\end{split}
\end{eqnarray}
\begin{eqnarray}\label{int_id2}
\begin{split}
& \int_0^{\infty} d v_\perp \int_{-\infty}^{\infty} d v_z h(v_\perp) l(v_z) P_z Q \\
=& \frac{1}{2\pi v_{\perp 0}} h(v_{\perp 0}) \int_0^{\infty} d v_z l(v_z) P_z \\
=& \frac{1}{2\pi v_{\perp 0}} h(v_{\perp 0}) [l(v_z) P(v_z)]_{-\infty}^{\infty} - \frac{1}{2\pi v_{\perp 0}} h(v_{\perp 0}) \int_{0}^{\infty} d v_z P \frac{d l}{d v_z} \\
=& -\frac{1}{2\pi v_{\perp 0}} h(v_{\perp 0}) \frac{d l}{d v_z}(u) .
\end{split}
\end{eqnarray}
Taking advantage of equation \eqref{int_id1} and \eqref{int_id2}, we can start to calculate each component of the susceptibility tensor. $\chi_{xx}$ can be calculated as
\begin{equation}
\chi_{xx} = - \frac{\omega_{pb}^2}{\omega^2} \sum\limits_{n=-\infty}^{\infty} \left[ \frac{2 n^2}{\lambda} J_n J_n^\prime + \frac{2 n^2}{\lambda} J_n J_n^\prime \frac{n \Omega_e}{\omega - k_z u - n\Omega_e} + J_n^2 \cot^2 \theta \frac{n^2 \Omega_e^2}{(\omega - k_z u - n \Omega_e)^2} \right] .
\end{equation}
Here $\lambda = k_\perp v_{\perp 0}/\Omega_e$. The argument of Bessel function and its derivatives is $\lambda$. $\theta$ is defined by $\tan \theta = k_\perp / k_z$. The first term can be further calculated as
\begin{eqnarray}\label{bessel_id1}
\begin{split}
\sum\limits_{n=-\infty}^{\infty} \frac{2 n^2}{\lambda} J_n J_n^\prime &= \sum\limits_{n=-\infty}^{\infty} \frac{n}{2} (J_{n-1}^2 - J_{n+1}^2) \\
&= \sum\limits_{n=-\infty}^{\infty} \left( \frac{n+1}{2} J_n^2 - \frac{n-1}{2} J_{n}^2 \right) \\
&= \sum\limits_{n=-\infty}^{\infty} J_n^2 \\
&= 1 .
\end{split}
\end{eqnarray}
Here the recurrence equation of Bessel function and its derivative are used. Thus
\begin{equation}
\chi_{xx} = - \frac{\omega_{pb}^2}{\omega^2} - \frac{\omega_{pb}^2}{\omega^2} \sum\limits_{n=-\infty}^{\infty} \left[ \left( \frac{2 n^2}{\lambda} J_n J_n^\prime \right) \frac{n \Omega_e}{\omega - k_z u - n\Omega_e} + \left( J_n^2 \cot^2 \theta \right) \frac{n^2 \Omega_e^2}{(\omega - k_z u - n \Omega_e)^2} \right] . 
\end{equation}
$\chi_{yy}$ can be calculated as
\begin{equation}
\chi_{yy} = - \frac{\omega_{pb}^2}{\omega^2} \sum\limits_{n=-\infty}^{\infty} \left[ \frac{1}{\lambda}(\lambda^2 (J_n^\prime)^2)^\prime + \frac{1}{\lambda}(\lambda^2 (J_n^\prime)^2)^\prime \frac{n \Omega_e}{\omega - k_z u - n\Omega_e} + \lambda^2 (J_n^\prime)^2 \cot^2 \theta \frac{\Omega_e^2}{(\omega - k_z u - n \Omega_e)^2} \right] .
\end{equation}
The first term can be further calculated as
\begin{eqnarray}
\begin{split}
\sum\limits_{n=-\infty}^{\infty} \frac{1}{\lambda}(\lambda^2 (J_n^\prime)^2)^\prime &= \sum\limits_{n=-\infty}^{\infty} \left[ 2 (J_n^\prime)^2 + 2 \lambda J_n^\prime J_n^{\prime \prime} \right] \\
&= \sum\limits_{n=-\infty}^{\infty} 2 J_n^\prime \left[ -\frac{1}{\lambda} (\lambda^2 - n^2) J_n \right] \\
&= -2 \lambda \sum\limits_{n=-\infty}^{\infty} J_n J_n^\prime + \sum\limits_{n=-\infty}^{\infty} \frac{2 n^2}{\lambda} J_n J_n^\prime \\
&= 1 .
\end{split}
\end{eqnarray}
Here we have used several identities, including the definition of the Bessel differential equation $\lambda^2 J_n^{\prime \prime} + \lambda J_n^\prime + (\lambda^2 - n^2) J_n = 0$, the identity $\sum\limits_{n=-\infty}^{\infty} J_n J_n^\prime = 0$ and the identity equation \eqref{bessel_id1}. Thus
\begin{equation}
\chi_{yy} = - \frac{\omega_{pb}^2}{\omega^2} - \frac{\omega_{pb}^2}{\omega^2} \sum\limits_{n=-\infty}^{\infty} \left[ \left( \frac{1}{\lambda}(\lambda^2 (J_n^\prime)^2)^\prime \right) \frac{n \Omega_e}{\omega - k_z u - n\Omega_e} + \left( \lambda^2 (J_n^\prime)^2 \cot^2 \theta \right) \frac{\Omega_e^2}{(\omega - k_z u - n \Omega_e)^2} \right] .
\end{equation}
$\chi_{zz}$ can be calculated as
\begin{equation}
\chi_{zz} = - \frac{\omega_{pb}^2}{\omega^2} \sum\limits_{n=-\infty}^{\infty} \left[ \left(\frac{2 n}{\lambda} J_n J_n^\prime \tan^2 \theta \right) \frac{k_z u}{\Omega_e} \frac{k_z u}{\omega - k_z u - n \Omega_e} + J_n^2 \frac{(\omega - n\Omega_e)^2}{(\omega - k_z u - n \Omega_e)^2} \right] .
\end{equation}
The first term in the summation can be further calculated as
\begin{equation}\label{bessel_id3}
\begin{split}
\sum\limits_{n=-\infty}^{\infty} \left(\frac{2 n}{\lambda} J_n J_n^\prime \right) \frac{k_z u}{\Omega_e} \frac{k_z u}{\omega - k_z u - n \Omega_e}
&= \sum\limits_{n=-\infty}^{\infty} \left(\frac{2 n}{\lambda} J_n J_n^\prime \right) \frac{k_z u}{\Omega_e} \left( -1 + \frac{\omega - n\Omega_e}{\omega - k_z u - n\Omega_e} \right) \\
&= \sum\limits_{n=-\infty}^{\infty} \left(\frac{2 n}{\lambda} J_n J_n^\prime \right) \frac{k_z u}{\Omega_e} \frac{\omega - n\Omega_e}{\omega - k_z u - n\Omega_e} \\
&= \sum\limits_{n=-\infty}^{\infty} \left(\frac{2 n}{\lambda} J_n J_n^\prime \right) \frac{\omega - n\Omega_e}{\Omega_e} \left( -1 + \frac{\omega - n\Omega_e}{\omega - k_z u - n\Omega_e} \right) \\
&= 1 + \sum\limits_{n=-\infty}^{\infty} \left(\frac{2 n}{\lambda} J_n J_n^\prime \right) \frac{(\omega - n \Omega_e)^2}{\Omega_e (\omega - k_z u - n \Omega_e)} .
\end{split}
\end{equation}
The second and fourth `=' sign in this summation use equation \eqref{bessel_id1} and the following identity
\begin{eqnarray}\label{bessel_id2}
\begin{split}
\sum\limits_{n=-\infty}^{\infty} \frac{2 n}{\lambda} J_n J_n^\prime &= \sum\limits_{n=-\infty}^{\infty} (J_{n-1} + J_{n+1}) \cdot \frac{1}{2}(J_{n-1} - J_{n+1}) \\
&= \frac{1}{2} \sum\limits_{n=-\infty}^{\infty} (J_{n-1}^2 - J_{n+1}^2) \\
&= 0 .
\end{split}
\end{eqnarray}
Thus
\begin{equation}
\chi_{zz} = - \frac{\omega_{pb}^2}{\omega^2} \tan^2 \theta - \frac{\omega_{pb}^2}{\omega^2} \sum\limits_{n=-\infty}^{\infty} \left[ \left(\frac{2 n}{\lambda} J_n J_n^\prime \tan^2 \theta \right) \frac{(\omega - n \Omega_e)^2}{\Omega_e (\omega - k_z u - n \Omega_e)} + J_n^2 \frac{(\omega - n\Omega_e)^2}{(\omega - k_z u - n \Omega_e)^2} \right] .
\end{equation}
$\chi_{xy}$ can be calculated as
\begin{equation}
\chi_{xy} = i \frac{\omega_{pb}^2}{\omega^2} \sum\limits_{n=-\infty}^{\infty} \left[ \frac{n}{\lambda} (\lambda J_n J_n^\prime)^\prime + \frac{n}{\lambda} (\lambda J_n J_n^\prime)^\prime \frac{n \Omega_e}{\omega - k_z u - n \Omega_e} + n \lambda J_n J_n^\prime \cot^2 \theta \frac{\Omega_e^2}{(\omega - k_z u - n \Omega_e)^2} \right] .
\end{equation}
The first term in the summation can be further calculated as
\begin{equation}\label{bessel_id4}
\begin{split}
\sum\limits_{n=-\infty}^{\infty} \frac{n}{\lambda} (\lambda J_n J_n^\prime)^\prime &= \sum\limits_{n=-\infty}^{\infty} n \left[ (J_n^\prime)^2 + J_n J_n^{\prime \prime} \right] \\
&= \sum\limits_{n=-\infty}^{\infty} \frac{n}{4} \left[ (J_{n-1} - J_{n+1})^2 + J_n (J_{n-2} - 2 J_n  + J_{n+2}) \right] \\
&= \frac{1}{4} \sum\limits_{n=-\infty}^{\infty} [ (n+1) J_n^2 - 2(n+1) J_n J_{n+2} \\
&\,\,\,\,\,\,\,+ (n-1) J_n^2 + (n+2) J_n J_{n+2} - 2n J_n^2 + n J_n J_{n+2} ] \\
&= 0 .
\end{split}
\end{equation}
Here the first `=' sign uses equation \eqref{bessel_id2}, the second `=' sign uses the recurrence relation and the third `=' sign changes the indexing of some terms. Thus
\begin{equation}
\chi_{xy} = i \frac{\omega_{pb}^2}{\omega^2} \sum\limits_{n=-\infty}^{\infty} \left[ \frac{n}{\lambda} (\lambda J_n J_n^\prime)^\prime \frac{n \Omega_e}{\omega - k_z u - n \Omega_e} + n \lambda J_n J_n^\prime \cot^2 \theta \frac{\Omega_e^2}{(\omega - k_z u - n \Omega_e)^2} \right] . 
\end{equation}
$\chi_{yx}$ can then be calculated as
\begin{eqnarray}
\chi_{yx} = - \chi_{xy} .
\end{eqnarray}
$\chi_{xz}$ can be calculated as
\begin{equation}
\chi_{xz} = - \frac{\omega_{pb}^2}{\omega^2} \sum\limits_{n=-\infty}^{\infty} \left[ -\frac{2 n^2}{\lambda} J_n J_n^\prime \tan \theta + \frac{2 n^2}{\lambda} J_n J_n^\prime \tan \theta \frac{\omega - n \Omega_e}{\omega - k_z u - n \Omega_e} + n J_n^2 \cot \theta \frac{\Omega_e (\omega - n\Omega_e)}{(\omega - k_z u - n \Omega_e)^2} \right] .
\end{equation}
Using equation \eqref{bessel_id1}, the first term in $\chi_xz$ can be simplified. Thus
\begin{equation}
\chi_{xz} = \frac{\omega_{pb}^2}{\omega^2} \tan \theta - \frac{\omega_{pb}^2}{\omega^2} \sum\limits_{n=-\infty}^{\infty} \left[ \left( \frac{2 n^2}{\lambda} J_n J_n^\prime \tan \theta \right) \frac{\omega - n \Omega_e}{\omega - k_z u - n \Omega_e} + \left( n J_n^2 \cot \theta \right) \frac{\Omega_e (\omega - n\Omega_e)}{(\omega - k_z u - n \Omega_e)^2} \right] .
\end{equation}
$\chi_{zx}$ can be calculated as
\begin{equation}
\chi_{zx} = - \frac{\omega_{pb}^2}{\omega^2} \sum\limits_{n=-\infty}^{\infty} \left[ \left( \frac{2 n}{\lambda} J_n J_n^\prime \tan\theta \right) \frac{k_z u}{\Omega_e} \frac{\omega - k_z u}{\omega - k_z u - n \Omega_e} + (n J_n^2 \cot\theta) \frac{\Omega_e (\omega - n\Omega_e)}{(\omega - k_z u - n \Omega_e)^2} \right] .
\end{equation}
The first term in the summation can be simplified using equation \eqref{bessel_id3}. Thus
\begin{equation}
\chi_{zx} = \frac{\omega_{pb}^2}{\omega^2} \tan\theta - \frac{\omega_{pb}^2}{\omega^2} \sum\limits_{n=-\infty}^{\infty} \left[ \left( \frac{2 n^2}{\lambda} J_n J_n^\prime \tan\theta \right) \frac{\omega - n \Omega_e}{\omega - k_z u - n \Omega_e} + (n J_n^2 \cot\theta) \frac{\Omega_e (\omega - n\Omega_e)}{(\omega - k_z u - n \Omega_e)^2} \right] .
\end{equation}
It is seen that
\begin{eqnarray}
\chi_{zx} = \chi_{xz} .
\end{eqnarray}
$\chi_{yz}$ can be calculated as
\begin{equation}
\resizebox{1.0\textwidth}{!}{ 
$\chi_{yz} = -i \frac{\omega_{pb}^2}{\omega^2} \sum\limits_{n=-\infty}^{\infty} \left[ - \frac{n}{\lambda} (\lambda J_n J_n^\prime)^\prime \tan\theta + \frac{n}{\lambda} (\lambda J_n J_n^\prime)^\prime \tan\theta \frac{\omega - n \Omega_e}{\omega - k_z v_z - n\Omega_e} + \lambda J_n J_n^\prime \cot\theta \frac{\Omega_e (\omega - n \Omega_e)}{(\omega - k_z u - n \Omega_e)^2} \right] . $
}
\end{equation}
The first term in this summation vanishes by using equation \eqref{bessel_id4}. Thus
\begin{equation}
\chi_{yz} = -i \frac{\omega_{pb}^2}{\omega^2} \sum\limits_{n=-\infty}^{\infty} \left[ \left( \frac{n}{\lambda} (\lambda J_n J_n^\prime)^\prime \tan\theta \right) \frac{\omega - n \Omega_e}{\omega - k_z v_z - n\Omega_e} + \left( \lambda J_n J_n^\prime \cot\theta \right) \frac{\Omega_e (\omega - n \Omega_e)}{(\omega - k_z u - n \Omega_e)^2} \right] .
\end{equation}
$\chi_{zy}$ can be calculated as
\begin{equation}
\chi_{zy} = i \frac{\omega_{pb}^2}{\omega^2} \sum\limits_{n=-\infty}^{\infty} \left[ \left( \frac{1}{\lambda} (\lambda J_n J_n^\prime)^\prime \tan\theta \right) \frac{\omega - k_z u}{\Omega_e} \frac{k_z u}{\omega - k_z u - n \Omega_e} + \left( \lambda J_n J_n^\prime \cot\theta \right) \frac{\Omega_e (\omega - n\Omega_e)}{(\omega - k_z u - n \Omega_e)^2} \right] .
\end{equation}
The first term in the summation can be further simplified as the following
\begin{equation}
\resizebox{.9\textwidth}{!}{ 
$\begin{split}
\sum\limits_{n=-\infty}^{\infty} \left( \frac{1}{\lambda} (\lambda J_n J_n^\prime)^\prime \right) \frac{\omega - k_z u}{\Omega_e} \frac{k_z u}{\omega - k_z u - n \Omega_e} &= \sum\limits_{n=-\infty}^{\infty} \left( \frac{1}{\lambda} (\lambda J_n J_n^\prime)^\prime \right) \frac{\omega - k_z u}{\Omega_e} \left( -1 + \frac{\omega - n \Omega_e}{\omega - k_z u - n \Omega_e} \right) \\
&= \sum\limits_{n=-\infty}^{\infty} \left( \frac{1}{\lambda} (\lambda J_n J_n^\prime)^\prime \right) \frac{\omega - k_z u}{\Omega_e} \frac{\omega - n \Omega_e}{\omega - k_z u - n \Omega_e} \\
&= \sum\limits_{n=-\infty}^{\infty} \left( \frac{1}{\lambda} (\lambda J_n J_n^\prime)^\prime \right) \frac{\omega - n \Omega_e}{\Omega_e} \left( 1 + \frac{n \Omega_e}{\omega - k_z u - n \Omega_e} \right) \\
&= \sum\limits_{n=-\infty}^{\infty} \left( \frac{n}{\lambda} (\lambda J_n J_n^\prime)^\prime \right) \frac{\omega - n \Omega_e}{\omega - k_z u - n \Omega_e} .
\end{split}$
}
\end{equation}
Here the second and fourth `=' sign use equation \eqref{bessel_id4} and the following identity
\begin{eqnarray}
\begin{split}
\sum\limits_{n=-\infty}^{\infty} \frac{1}{\lambda} (\lambda J_n J_n^\prime)^\prime &= \sum\limits_{n=-\infty}^{\infty} \frac{1}{\lambda} (J_n J_n^\prime + \lambda (J_n J_n^\prime)^\prime) \\
&= \sum\limits_{n=-\infty}^{\infty} (J_n J_n^\prime)^\prime \\
&= \sum\limits_{n=-\infty}^{\infty} \left( (J_n^\prime)^2 + J_n J_n^{\prime \prime} \right) \\
&= \frac{1}{4} \sum\limits_{n=-\infty}^{\infty} \left[ (J_{n-1} - J_{n+1})^2 + J_n (J_{n-2} - 2 J_n + J_{n+2}) \right] \\
&= 0 ,
\end{split}
\end{eqnarray}
where we have used $\sum\limits_{n=-\infty}^{\infty} J_n J_n^\prime = 0$, the recurrence relation of Bessel function and the change of indexing technique. Thus $\chi_{zy}$ can be rewritten as
\begin{equation}
\chi_{zy} = i \frac{\omega_{pb}^2}{\omega^2} \sum\limits_{n=-\infty}^{\infty} \left[ \left( \frac{n}{\lambda} (\lambda J_n J_n^\prime)^\prime \tan\theta \right) \frac{\omega - n \Omega_e}{\omega - k_z u - n \Omega_e} + \left( \lambda J_n J_n^\prime \cot\theta \right) \frac{\Omega_e (\omega - n\Omega_e)}{(\omega - k_z u - n \Omega_e)^2} \right] .
\end{equation}
It is seen that
\begin{eqnarray}
\chi_{zy} = - \chi_{yz} .
\end{eqnarray}
To summarize the results, we collect all the components of the susceptibility tensor as the following
\begin{equation}
\chi_{xx} = - \frac{\omega_{pb}^2}{\omega^2} - \frac{\omega_{pb}^2}{\omega^2} \sum\limits_{n=-\infty}^{\infty} \left[ \left( \frac{2 n^2}{\lambda} J_n J_n^\prime \right) \frac{n \Omega_e}{\omega - k_z u - n\Omega_e} + \left( J_n^2 \cot^2 \theta \right) \frac{n^2 \Omega_e^2}{(\omega - k_z u - n \Omega_e)^2} \right] ,
\end{equation}
\begin{equation}
\chi_{yy} = - \frac{\omega_{pb}^2}{\omega^2} - \frac{\omega_{pb}^2}{\omega^2} \sum\limits_{n=-\infty}^{\infty} \left[ \left( \frac{1}{\lambda}(\lambda^2 (J_n^\prime)^2)^\prime \right) \frac{n \Omega_e}{\omega - k_z u - n\Omega_e} + \left( \lambda^2 (J_n^\prime)^2 \cot^2 \theta \right) \frac{\Omega_e^2}{(\omega - k_z u - n \Omega_e)^2} \right] ,
\end{equation}
\begin{equation}
\chi_{zz} = - \frac{\omega_{pb}^2}{\omega^2} \tan^2 \theta - \frac{\omega_{pb}^2}{\omega^2} \sum\limits_{n=-\infty}^{\infty} \left[ \left(\frac{2 n}{\lambda} J_n J_n^\prime \tan^2 \theta \right) \frac{(\omega - n \Omega_e)^2}{\Omega_e (\omega - k_z u - n \Omega_e)} + J_n^2 \frac{(\omega - n\Omega_e)^2}{(\omega - k_z u - n \Omega_e)^2} \right] ,
\end{equation}
\begin{equation}
\chi_{xy} = i \frac{\omega_{pb}^2}{\omega^2} \sum\limits_{n=-\infty}^{\infty} \left[ \left( \frac{n}{\lambda} (\lambda J_n J_n^\prime)^\prime \right) \frac{n \Omega_e}{\omega - k_z u - n \Omega_e} + \left( n \lambda J_n J_n^\prime \cot^2 \theta \right) \frac{\Omega_e^2}{(\omega - k_z u - n \Omega_e)^2} \right] ,
\end{equation}
\begin{equation}
\chi_{xz} = \frac{\omega_{pb}^2}{\omega^2} \tan \theta - \frac{\omega_{pb}^2}{\omega^2} \sum\limits_{n=-\infty}^{\infty} \left[ \left( \frac{2 n^2}{\lambda} J_n J_n^\prime \tan \theta \right) \frac{\omega - n \Omega_e}{\omega - k_z u - n \Omega_e} + \left( n J_n^2 \cot \theta \right) \frac{\Omega_e (\omega - n\Omega_e)}{(\omega - k_z u - n \Omega_e)^2} \right] ,
\end{equation}

\begin{equation}
\chi_{yz} = -i \frac{\omega_{pb}^2}{\omega^2} \sum\limits_{n=-\infty}^{\infty} \left[ \left( \frac{n}{\lambda} (\lambda J_n J_n^\prime)^\prime \tan\theta \right) \frac{\omega - n \Omega_e}{\omega - k_z v_z - n\Omega_e} + \left( \lambda J_n J_n^\prime \cot\theta \right) \frac{\Omega_e (\omega - n \Omega_e)}{(\omega - k_z u - n \Omega_e)^2} \right] ,
\end{equation}
\begin{equation}
\chi_{yx} = - \chi_{xy} ,
\end{equation}
\begin{equation}
\chi_{zx} = \chi_{xz} ,
\end{equation}
\begin{equation}
\chi_{zy} = - \chi_{yz} .
\end{equation}

\end{document}